\documentclass{aa}

\usepackage[varg]{txfonts}
\usepackage[colorlinks=true, linkcolor=blue, citecolor=blue, urlcolor=gray]{hyperref}
\usepackage{xcolor}
\usepackage{listings}
\usepackage{algpseudocode}

\bibpunct{(}{)}{;}{a}{}{,}

\newcommand{\txd}{{\text{d}}}

\newcommand{\bfx}{{\bf{x}}}

\DeclareMathOperator{\sign}{sign}

\begin{document}

\title{Mass-conservative voxelisation of smoothed particles \\
using cumulative kernels}

\titlerunning{Mass-conservative voxelisation}

\author{%
Maarten Baes\inst{\ref{UGent}}\corrauth{maarten.baes@ugent.be} 
\and
Peter Camps\inst{\ref{UGent}}\email{peter.camps@ugent.be}
\and
Kosei Matsumoto\inst{\ref{UGent}}\email{kosei.matsumoto@ugent.be}
\and
Jules Dierickx\inst{\ref{UGent}}\email{jules.dierickx@ugent.be}
\and
\\
Andrea Gebek\inst{\ref{UGent},\ref{Bath}}\email{andrea.gebek@ugent.be}
\and
Angelos Nersesian\inst{\ref{Liege},\ref{UGent}}\email{angelos.nersesian@ugent.be}
}

\institute{%
Department of Physics and Astronomy, Universiteit Gent, Proeftuinstraat 86 N3, B-9000 Ghent, Belgium
\label{UGent}
\and
Department of Physics, University of Bath, Claverton Down, Bath, BA2 7AY, UK
\label{Bath}
\and
STAR Institute, Universit\'e de Li\`ege, Quartier Agora, All\'ee du Six Ao\^ut 19c, B-4000 Li\`ege, Belgium
\label{Liege}
}

\authorrunning{M.~Baes et al.}

\date{\today}

\abstract 
{Smoothed particle hydrodynamics (SPH) simulations are frequently post-processed on Cartesian grids for applications such as radiative transfer calculations, visualisation, and data analysis. While numerous voxelisation algorithms have been proposed, relatively little attention has been paid to the fundamental properties that such algorithms should satisfy.}
{We argue that SPH voxelisation algorithms should accurately reproduce the underlying density field, globally conserve mass, remain consistent under grid refinement, and be computationally efficient. We present a new finite-volume algorithm designed to satisfy these requirements simultaneously.}
{Starting from a finite-volume formulation of the SPH density field, we derived a simple expression for the mass contained in a Cartesian voxel in terms of a cumulative kernel function. The cumulative kernel was tabulated numerically once for any compact-support smoothing kernel, independently of its analytical form, so that the same tabulation and voxelisation machinery applies unchanged to any such kernel. During voxelisation, mass contributions were obtained from interpolated evaluations of this precomputed function at voxel corners, with values at corners shared by neighbouring voxels evaluated only once and reused. Using direct high-order numerical integration as a reference for the voxel densities, we compared the new cumulative-kernel algorithm to nearest-grid-point assignment, centre sampling, regular subsampling, and Monte Carlo integration and implemented it in the 3D Monte Carlo radiative transfer code {\texttt{SKIRT}}.}
{The cumulative-kernel algorithm reproduces the reference voxel densities substantially more accurately than the sampling-based methods considered. For the adopted $129^3$ cumulative-kernel lookup table, the median absolute relative density error is approximately $4\times10^{-5}$, and the error decreases approximately quadratically with increasing lookup-table resolution. The method conserves mass and remains consistent under arbitrary grid refinement approximately to machine precision. It is substantially more efficient than the sampling-based methods considered, which account for the finite extent of the smoothing kernel. Its implementation in {\texttt{SKIRT}} applies to both Cartesian and hierarchical octree and binary-tree grids and accelerates cell-density calculations by approximately a factor of five and hierarchical tree construction by approximately a factor of four compared to the default Monte Carlo approach.}
{The cumulative-kernel voxelisation algorithm provides an accurate, mass-conservative, refinement-consistent, and computationally efficient method for voxelising SPH particle distributions. The method is independent of the analytical form of the smoothing kernel and can therefore be applied straightforwardly to a wide range of SPH kernels. We advocate its use in radiative transfer post-processing and other voxelisation applications.}


\maketitle
\nolinenumbers

\section{Introduction}
\label{Introduction.sec}

Smoothed particle hydrodynamics (SPH) is one of the standard techniques for simulating astrophysical fluids \citep{Monaghan1992, Springel2010b, Price2012}. Owing to its Lagrangian nature, exact mass conservation, and natural adaptivity to large density contrasts, SPH has found widespread application in computational astrophysics, including simulations of star formation, galaxy formation and evolution, cosmological structure formation, and compact-object mergers. A variety of mature SPH codes are now routinely used by the community, including {\texttt{Gasoline}} \citep{Wadsley2004}, {\texttt{GADGET}} \citep{Springel2005, Springel2021}, {\texttt{Phantom}} \citep{Price2018}, and {\texttt{SWIFT}} \citep{Schaller2024}. In SPH, the fluid is represented by a set of particles, each carrying mass, thermodynamic properties, and a smoothing length, from which a continuous density field can be reconstructed through a smoothing kernel.

While this particle-based representation is ideally suited for hydrodynamical simulations, many post-processing applications require the density field to be represented on a spatial grid. One of the most important examples is radiative transfer post-processing, in which the simulated 3D density distribution is used to generate synthetic images, spectra, and data cubes that can be compared directly to observations \citep{Steinacker2013, Noebauer2019}. In most 3D radiative transfer codes, the medium is discretised on a grid. Although several grid types can be employed for this purpose, including unstructured Voronoi tessellations \citep{Camps2013, Hubber2016} or tetrahedral grids \citep{Lauwers2024}, Cartesian grids, either regular or hierarchically refined, remain the most widely used choice because of their conceptual simplicity and computational efficiency \citep[e.g.][]{Kurosawa2001, Wolf2003, Bianchi2008, Saftly2013, Saftly2014}.

Representations of SPH particles on 3D Cartesian grids, commonly referred to as voxelisations, are also widely used for the visualisation and analysis of astrophysical SPH simulations. Software packages in which such algorithms are implemented include {\texttt{SPLASH}} \citep{Price2007}, {\texttt{Sarracen}} \citep{Harris2023}, {\texttt{Py-SPHViewer}} \citep{BenitezLlambay2017}, and {\texttt{SWIFTsimIO}} \citep{Borrow2020}. Outside astrophysics, similar voxelisation techniques are required in related fields such as computer graphics or fluid dynamics \citep[e.g.][]{Zhang2008, Cha2009, Fraedrich2010, Shi2013, Zirr2018, Duque2020, Piochowiak2021, Rosswog2021}.

Despite the widespread use of voxelisation, the fundamental properties that voxelisation algorithms should satisfy have received comparatively little attention. We argue that a suitable voxelisation algorithm should meet four key criteria. First and foremost, it should accurately represent the continuous SPH density field or equivalently its finite-volume averages on the adopted grid. Second, it should be globally mass-conservative, such that the total mass represented on the grid equals the total particle mass. A stronger requirement is consistency under grid refinement: refining the computational grid should merely redistribute the mass over smaller voxels without changing the mass assigned to any finite region. Finally, the algorithm should be computationally efficient, remaining practical for simulations containing millions of particles and millions of grid cells.

Existing voxelisation algorithms can broadly be divided into three classes according to the underlying philosophy. The first category consists of particle assignment schemes, such as nearest-grid-point (NGP) or cloud-in-cell (CIC) interpolation \citep[e.g.][]{Hockney1981}. These methods, widely used in $N$-body simulations, assign the mass of each particle to one or more neighbouring grid cells according to a prescribed interpolation rule. They are exactly mass-conservative and computationally efficient, but they ignore the smoothing kernel associated with the SPH particles and therefore do not accurately represent the continuous SPH density field.

A second category consists of density sampling methods, which evaluate the SPH density field at one or more locations within each voxel. The simplest approach samples the density at the voxel centre, while more accurate variants estimate the voxel average through deterministic subsampling or Monte Carlo integration \citep[e.g.][]{Kurosawa2001, Koepferl2017}. Increasing the number of sampling points improves the accuracy of the estimated voxel average, but these methods generally remain neither exactly mass-conservative nor consistent under grid refinement, while their computational cost increases rapidly with the number of sampling points. In some cases, density sampling methods are combined with particle assignment schemes, depending on the ratio of the smoothing length to the voxel scale \citep{Dolag2005, Rottgers2018, BenitezLlambay2025}.

Finally, finite-volume methods determine the particle contribution to each voxel by integrating the smoothing kernel over the voxel volume. These methods satisfy the mass-conservation and refinement-consistency requirements by construction. For Gaussian smoothing kernels, this integration is particularly straightforward because the kernel is separable in the three Cartesian coordinates \citep{Altay2013}. This property was exploited by \citet{Camps2016b} to calculate the mass within Cartesian grid cells analytically as a product of error-function differences. This separability, however, is an exact property only of the infinite, untruncated Gaussian. Once the smoothing kernel is truncated to enforce compact support, as commonly implemented in SPH codes, the truncation condition couples the three coordinates, and the analytic separable solution is no longer exact but only an approximation to the true, truncated kernel.

A more general formulation is presented by \citet{Petkova2018}, who derived an exact analytical solution for arbitrary convex polyhedral cells. While elegant and highly general, this formulation requires rather involved and kernel-specific analytical expressions and geometric bookkeeping, making it more complex than is necessary for the common case of Cartesian voxelisations.

In this paper we present a new finite-volume voxelisation algorithm for Cartesian grids. The central idea is to replace kernel-specific analytical antiderivatives by a numerically tabulated cumulative kernel. Conceptually, the proposed method separates the kernel-specific part of the calculation, which is performed only once during the construction of the cumulative kernel, from voxelisation itself, which thereafter becomes completely independent of the analytical form of the smoothing kernel. By evaluating the cumulative kernel at voxel corners and reusing values shared by neighbouring voxels, voxelisation satisfies all four criteria identified above: it accurately represents the SPH density field, is globally mass-conserving and consistent under grid refinement, and remains computationally efficient.

The remainder of this paper is organised as follows. In Sect.~{\ref{VoxelisationDescription.sec}}, we formulate the voxelisation problem and derive the cumulative-kernel finite-volume algorithm. Section~{\ref{Implementation.sec}} describes its practical implementation and the comparison methods used in this work. In Sect.~{\ref{Validation.sec}}, we validate the algorithm by comparing it to commonly used voxelisation methods in terms of accuracy of the voxelised density field, global mass conservation, consistency under grid refinement, and computational efficiency. In Sect.~{\ref{SKIRT.sec}}, we demonstrate the applicability of the method in a realistic production environment through its implementation in the {\texttt{SKIRT}} radiative transfer code. Finally, Sect.~{\ref{Conclusions.sec}} summarises our main conclusions.

\section{Voxelisation of smoothed particles}
\label{VoxelisationDescription.sec}

\subsection{General problem description}

Let us assume a medium described by a list of $N_{\text{part}}$ smoothed particles. Each particle $n$ is characterised by its position $\bfx_n$, its smoothing length $h_n$, and its mass $m_n$. A smoothing kernel $W(u)$ describes the spatial distribution of the mass associated with each particle. Following the convention\footnote{This convention differs from the one used by some SPH codes, including {\texttt{Gadget}}, {\texttt{Swift}}, and {\texttt{Phantom}}, in which the tabulated smoothing length corresponds to compact support at $r/h \leqslant 2$. Users importing particle data from such codes should multiply the reported smoothing lengths by a factor of two before applying the algorithm described here, or equivalently rescale the tabulated kernel, to avoid a spurious error by a factor of two in the effective kernel radius.} used in some SPH codes and in the {\texttt{SKIRT}} radiative transfer code, we assume throughout that the kernel is normalised and has compact support over $0\leqslant u \leqslant1$, such that $W(u)=0$ for $u>1$. Given the list of particles and the choice of the smoothing kernel, the spatial density at any position $\bfx$ is 
\begin{equation}
\rho(\bfx) = \sum_{n=1}^{N_{\text{part}}} \frac{m_n}{h_n^3}\,W(u_n),
\label{rho}
\end{equation}
where we define
\begin{equation}
u_n = u_n(\bfx) = \frac{\| \bfx - \bfx_n \|}{h_n}.
\end{equation}
Suppose we wish to represent this density field on a 3D Cartesian grid consisting of $N_{\text{vox}}$ cuboidal voxels with centres $\bfx_i$, i.e.\ we seek an algorithm that determines the value $\rho_i$ for each cell. As discussed in Sect.~\ref{Introduction.sec}, this algorithm should accurately represent the SPH density field, be globally mass-conservative, be consistent under grid refinement, and be computationally efficient. The mass-conservation criterion implies that 
\begin{equation}
\sum_{n=1}^{N_{\text{part}}} m_n = \sum_{i=1}^{N_{\text{vox}}} \rho_i V_i \equiv \sum_{i=1}^{N_{\text{vox}}} M_i,
\end{equation}
with $V_i$ and $M_i$ being the volume and the mass of cell number $i$, respectively. The refinement-consistency criterion can be written as
\begin{equation}
M_i = \sum_{j=1}^{N_{\text{ss}}} M_{i,j}
\end{equation}
for any partition of cell number $i$ into $N_{\text{ss}}$ subcells.

Sampling methods estimate the voxel density from one or more evaluations of the continuous density field (see Sect.~\ref{Introduction.sec}). While these methods can provide increasingly accurate approximations as the number of sampling positions increases, they do not satisfy the mass-conservation and refinement-consistency criteria exactly. In contrast, we seek a finite-volume formulation in which the voxel mass is obtained by integrating the density over the voxel volume,
\begin{equation}
M_i = \int_{V_i} \rho(\bfx)\,\txd\bfx,
\end{equation}
or explicitly for the case of smoothed particles,
\begin{equation}
M_i = \sum_{n=1}^{N_{\text{part}}} \frac{m_n}{h_n^3} \int_{V_i} W(u_n)\,\txd\bfx.
\label{Mi}
\end{equation}
The corresponding voxel density is then obtained as
\begin{equation}
\rho_i = \frac{M_i}{V_i}.
\end{equation}
The key to a mass-conservative voxelisation algorithm is a method that exactly and efficiently evaluates expression~(\ref{Mi}).

\subsection{Finite-volume voxelisation using a cumulative kernel}

\citet{Petkova2018} present an elegant method for evaluating expression (\ref{Mi}). They demonstrate that this volume integral can be evaluated exactly by applying Gauss' divergence theorem to transform it into a surface integral over the cell boundary. A subsequent application of Stokes' theorem reduces the surface integrals to line integrals along the edges of the polyhedron. The resulting line integrals depend only on the vertices and edges of the polyhedral cell, which makes the method applicable to arbitrary convex polyhedra. For the specific case of the cubic spline kernel, these line integrals can be evaluated analytically, yielding an exact algorithm that is applicable to arbitrary convex polyhedral cells, including Cartesian voxels and Voronoi cells. 

The formulation of \citet{Petkova2018} is mathematically elegant, allowing the exact evaluation of the mass contribution of smoothed particles to arbitrary convex polyhedral cells. However, many applications require voxelisation only onto regular Cartesian grids. In this more restricted setting, the full geometric generality of this formulation is unnecessary. By exploiting the regular geometry of Cartesian voxels, the particle weights can be evaluated using a substantially simpler formulation.

For our algorithm, we assume that the Cartesian cell with label $i$ has boundaries
$x_i^- < x <  x_i^+$, $y_i^- < y <  y_i^+$, and $z_i^- < z <  z_i^+$. The mass in cell number $i$ is then
\begin{equation}
M_i 
=
\sum_{n=1}^{N_{\text{part}}} 
\frac{m_n}{h_n^3}
\int_{x_i^-}^{x_i^+} \txd x
\int_{y_i^-}^{y_i^+} \txd y
\int_{z_i^-}^{z_i^+} W(u_n)\, \txd z.
\label{Mijk-recast}
\end{equation}

Evaluating the triple integral in Eq.~(\ref{Mijk-recast}) directly for every particle--voxel pair would be both cumbersome and inefficient, since the integrand depends on the kernel's analytical form and the integration limits differ for every voxel. We simplified this problem in three steps: we rescaled coordinates by each particle's smoothing length to obtain a fixed integrand; we defined the triple antiderivative of the kernel as a cumulative kernel $\Phi$, tabulated once for the adopted kernel; and we expressed the voxel mass as a signed sum of eight corner evaluations of $\Phi$, the 3D analogue of the Newton--Leibniz rule. The remainder of this subsection works through these three steps in turn.

For each value of $n$, we can transform the integration variables in Eq.~(\ref{Mijk-recast}) to 
\begin{gather}
X = \frac{x-x_n}{h_n}, \quad
Y = \frac{y-y_n}{h_n}, \quad
Z = \frac{z-z_n}{h_n},
\end{gather}
which results in 
\begin{equation}
M_i 
=
\sum_{n=1}^{N_{\text{part}}} 
m_n
\int_{X_{i,n}^-}^{X_{i,n}^+} \txd X
\int_{Y_{i,n}^-}^{Y_{i,n}^+} \txd Y
\int_{Z_{i,n}^-}^{Z_{i,n}^+}  W\left(\sqrt{X^2+Y^2+Z^2}\right) \txd Z,
\end{equation}
with 
\begin{gather}
X_{i,n}^\pm = \frac{x_i^\pm - x_n}{h_n},
\quad
Y_{i,n}^\pm = \frac{y_i^\pm - y_n}{h_n},
\quad
Z_{i,n}^\pm = \frac{z_i^\pm - z_n}{h_n}.
\label{corners}
\end{gather}
This simple transformation turns the expression for $M_i$ into a sum of triple integrals with exactly the same integrand but with a different rectangular integration domain for each particle. We now define the cumulative kernel
\begin{equation}
\Phi(X,Y,Z) \equiv \int_0^X \txd X' \int_0^Y \txd Y' \int_0^Z W\left(\sqrt{X'^2+Y'^2+Z'^2}\right)\txd Z'\,,
\label{defW3}
\end{equation}
for $X\geqslant0$, $Y\geqslant0$ and $Z\geqslant0$. Because the kernel is symmetric in $X'$, $Y'$ and $Z'$, we define the signed primitive
\begin{equation}
\Phi^*(X,Y,Z) = \sign X \sign Y \sign Z\, \Phi(|X|,|Y|,|Z|).
\end{equation}
The exact voxel contribution is now a linear combination of eight evaluations of the trivariate function $\Phi^*(X,Y,Z)$. Using simple geometry we find  
\begin{equation}
M_i = \sum_{n=1}^{N_{\text{part}}} m_n
\sum_{\alpha,\beta,\gamma \in \{-,+\}}
(-1)^{N_-(\alpha,\beta,\gamma)}\,\Phi^*\Bigl(X_{i,n}^\alpha, Y_{i,n}^\beta, Z_{i,n}^\gamma\Bigr), 
\label{Mifinal}
\end{equation}
where $N_-$ is the number of minus signs in the triple $(\alpha,\beta,\gamma)$. Explicitly we have 
\begin{align}
M_i = \sum_{n=1}^{N_{\text{part}}} m_n
\,\biggl[
&
\Phi^*\Bigl(X_{i,n}^+, Y_{i,n}^+, Z_{i,n}^+\Bigr) 
- \Phi^*\Bigl(X_{i,n}^+, Y_{i,n}^+, Z_{i,n}^-\Bigr) 
\nonumber \\[-1ex]
&\quad
- \Phi^*\Bigl(X_{i,n}^+, Y_{i,n}^-, Z_{i,n}^+\Bigr) 
+ \Phi^*\Bigl(X_{i,n}^+, Y_{i,n}^-, Z_{i,n}^-\Bigr) 
\nonumber \\[0.4ex]
&\quad
- \Phi^*\Bigl(X_{i,n}^-, Y_{i,n}^+, Z_{i,n}^+\Bigr) 
+ \Phi^*\Bigl(X_{i,n}^-, Y_{i,n}^+, Z_{i,n}^-\Bigr) 
\nonumber \\
&\quad
+ \Phi^*\Bigl(X_{i,n}^-, Y_{i,n}^-, Z_{i,n}^+\Bigr) 
- \Phi^*\Bigl(X_{i,n}^-, Y_{i,n}^-, Z_{i,n}^-\Bigr) 
\,\biggr].
\label{Mifinal2}
\end{align}
As anticipated above, Eqs.~(\ref{Mifinal}) and (\ref{Mifinal2}) are the 3D analogues of the Newton--Leibniz theorem. Just as a 1D definite integral is obtained from the difference of two primitive values, the mass contained in a voxel is obtained as the triple finite difference of the cumulative kernel.

Unlike the analytical approach of \citet{Petkova2018}, the cumulative-kernel method does not require a closed-form expression for $\Phi$. Instead, the cumulative kernel $\Phi$ can be computed numerically once and stored in a lookup table. Once this lookup table has been constructed, the contribution to each cell is obtained from eight interpolated cumulative-kernel values at its corners.

\section{Implementation}
\label{Implementation.sec}

\subsection{Construction of the cumulative kernel}

The derivation of the voxelisation algorithm is entirely independent of the analytical form of the smoothing kernel. The only kernel-dependent quantity is the cumulative kernel $\Phi(X,Y,Z)$, which can therefore be tabulated once for any compact-support kernel and subsequently reused for an arbitrary number of voxelisation calculations.

For a given smoothing kernel, the cumulative kernel is tabulated on a regular Cartesian grid covering the positive octant $0\leqslant X,Y,Z \leqslant 1$. The value of $\Phi$ at every grid point is obtained through direct numerical evaluation of Eq.~(\ref{defW3}) using the 64-point Gauss--Legendre integrator borrowed from the {\texttt{SpheCow}} code \citep{Baes2021b}. Since $\Phi$ is obtained by three successive integrations of the smoothing kernel, it is a very smooth function. Consequently, relatively modest lookup tables together with simple trilinear interpolation already provide excellent accuracy. The resulting 3D lookup tables are stored in FITS format and read during the voxelisation procedure.

\subsection{Voxelisation algorithm}
\label{Algorithm.sec}

During voxelisation, particles are processed one by one. For each particle, only those voxels that overlap the compact support of the smoothing kernel need to be considered. Since all kernels considered in this work have support $u\leqslant1$, this corresponds to the rectangular region
\begin{equation}
x_n - h_n \leqslant x \leqslant x_n + h_n,
\end{equation}
and similarly for the $y$- and $z$-coordinates. Particles whose support extends beyond the computational domain are handled by restricting the voxel iteration to the voxels inside the grid. The normalised coordinates of all distinct corner points associated with the overlapping voxels are calculated according to Eqs.~(\ref{corners}), and the cumulative kernel is evaluated at these points through trilinear interpolation within the lookup table. The resulting values are stored temporarily. For each overlapping voxel, the eight corresponding cumulative-kernel values are then retrieved, and the voxel mass contribution follows directly from Eq.~(\ref{Mifinal}).

Storing the cumulative-kernel values at the voxel corners provides an important optimisation, since neighbouring voxels share corner points. A straightforward implementation of Eq.~(\ref{Mifinal}) would require eight cumulative-kernel evaluations for every particle--voxel overlap and would therefore repeatedly evaluate the same values. If a particle overlaps $N_x\times N_y\times N_z$ voxels, such an implementation requires eight $\,N_xN_yN_z$ cumulative-kernel evaluations, whereas the stored-corner implementation requires only $(N_x+1)(N_y+1)(N_z+1)$. For particles whose smoothing volumes cover many voxels, the number of evaluations is thus reduced by almost a factor of eight. The additional memory requirement is modest, since the stored values only need to be retained while processing a single particle.

Pseudocode for the entire algorithm is
\vspace*{2ex}
\begin{algorithmic}
\it
\State initialise voxel masses to zero
\For{each particle}
   \State determine voxels intersecting the kernel support
   \State determine all voxel corners associated with these voxels
   \State evaluate and store the cumulative kernel at each corner
   \For{each intersecting voxel}
      \State retrieve the eight cumulative-kernel values
      \State determine particle contribution using Eq.~(14)
      \State accumulate voxel mass
   \EndFor
\EndFor
\State divide voxel masses by voxel volumes.
\end{algorithmic}
\vspace*{2ex}

A C++ version of the implementation can be found on GitHub\footnote{\url{https://github.com/mbaes/Voxelise}}. This software distribution also contains pre-generated FITS lookup tables for several common smoothing kernels, as well as a utility program to generate them for any other kernel.

We note that the order of the iteration can also be reversed. Instead of processing the particles one by one and identifying the voxels overlapping each particle, one can iterate over the voxels and identify the particles whose smoothing volumes overlap each voxel. For the regular Cartesian grids considered here, the particle-oriented approach is more efficient because the range of potentially overlapping voxels can be determined directly from the particle position and smoothing length. The cell-oriented approach can, however, be more convenient for more general grid structures. A disadvantage is that cumulative-kernel values at corners shared by neighbouring cells cannot readily be reused, as they can in the particle-oriented implementation described above. An example of such a cell-oriented implementation for Cartesian and hierarchical grids is discussed in Sect.~{\ref{SKIRT.sec}}.

\begin{figure*}
\sidecaption
\includegraphics[width=0.7\textwidth]{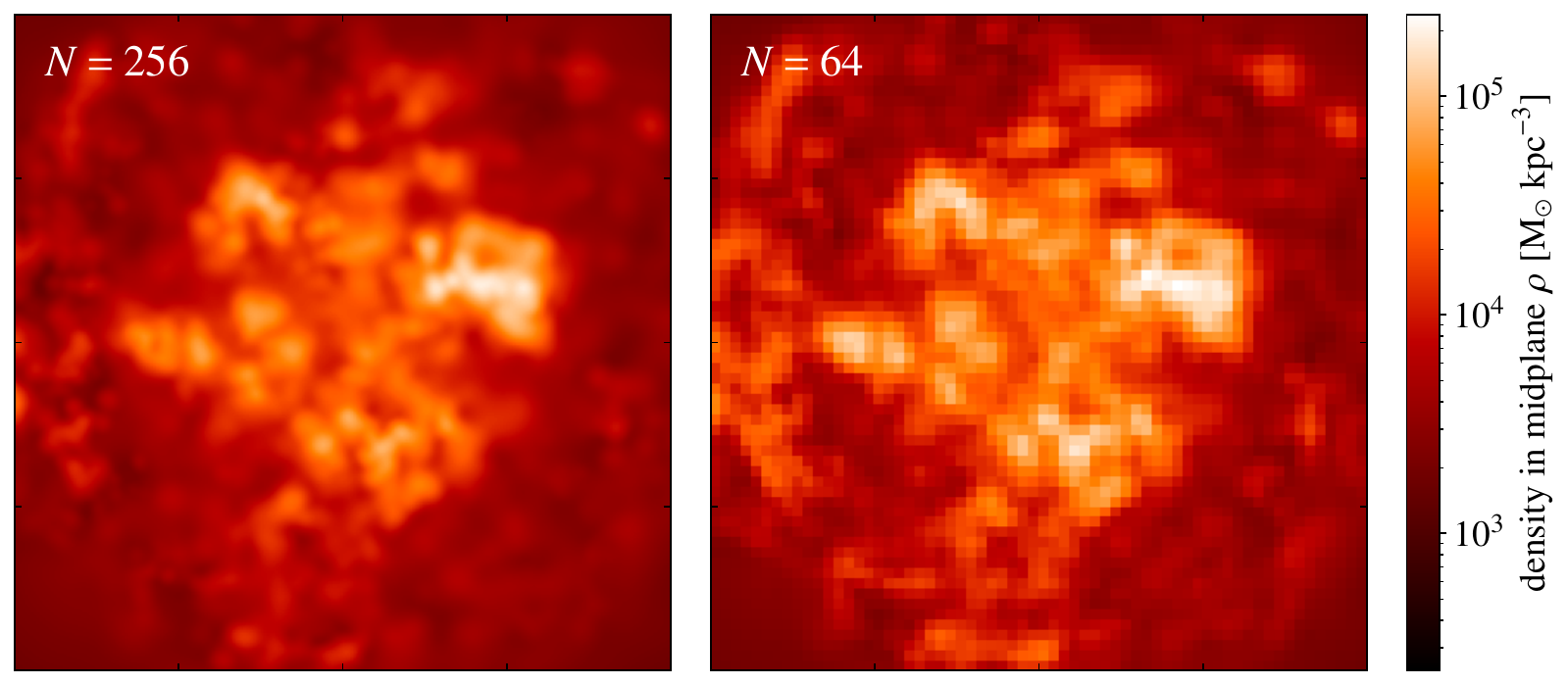}
\caption{Midplane density distribution of the simulated galaxy used for the numerical validation. The panels show the central plane of cumulative-kernel voxelisations with linear grid resolutions $N=256$ (left) and $N=64$ (right). Owing to the orientation of the galaxy in the simulation volume, this plane approximately corresponds to the galactic midplane. The $64^3$ grid is used for the density-accuracy comparison in Figs.~{\ref{NGC628_relerr.fig}} and~{\ref{NGC628_histograms.fig}}.}
\label{NGC628_density.fig}
\end{figure*}

\subsection{Comparison methods}
\label{ComparisonMethods.sec}

To assess the performance of the cumulative-kernel voxelisation algorithm, we compared it to a number of alternative voxelisation methods that are commonly used in practice. All methods were implemented within the same C++ framework and used identical particle input, grid construction, voxel overlap selection, and output routines. The only difference between the various implementations was the procedure used to assign particle mass to the voxels. This ensured that the measured differences in accuracy and runtime reflected only the voxelisation algorithm itself.

The first comparison method was the NGP algorithm, in which the entire mass of each particle is assigned to the voxel containing the particle centre. This method ignores the smoothing kernel entirely and therefore does not reconstruct the SPH density field. Nevertheless, because every particle contributes its full mass to exactly one voxel, the method is exactly mass-conservative and consistent under grid refinement.

The other comparison methods we considered were sampling methods. The simplest was centre sampling, in which we approximated the density associated with a voxel by evaluating Eq.~(\ref{rho}) at the voxel centre. As a first improvement over centre sampling, we considered regular subsampling. In this approach, the voxel density is estimated by averaging the kernel values at the central positions of a regular Cartesian subgrid within each voxel. We considered two sampling resolutions, consisting of $3\times3\times3$ and $5\times5\times5$ sampling points per voxel. Finally, we considered Monte Carlo sampling, in which the voxel density is estimated from kernel evaluations at randomly distributed positions within the voxel. We used 10, 100, and 1000 random sampling points per voxel. 

\section{Numerical validation}
\label{Validation.sec}

Having introduced the four requirements a voxelisation algorithm should satisfy (accurate reconstruction of the density field, global mass conservation, consistency under grid refinement, and computational efficiency), we now assess the cumulative-kernel algorithm against each of these requirements in turn. Throughout, we compare its performance with the alternative methods described in Sect.~{\ref{ComparisonMethods.sec}}.

\subsection{Test setup}
\label{NGC628.sec}

As a representative test case, we considered the gas distribution in an SPH simulation of an isolated disc galaxy. The simulation was run with the {\texttt{GADGET4-Osaka}} code \citep{Romano2022a, Romano2022b}, which builds on the {\texttt{GADGET-4}} SPH engine \citep{Springel2021}. It was designed to resemble NGC\,628, one of the nearest and most thoroughly studied face-on disc galaxies. A detailed analysis of the simulation and the comparison to NGC\,628 was presented by \citet{vanderGiessen2024} and \citet{Matsumoto2024}. The gas distribution consisted of 399,154 individual SPH particles.

For the numerical experiments presented in this section, we used the cubic spline kernel \citep{Monaghan1985}. With the normalisation condition adopted in this work, this kernel is given by
\begin{equation}
W(u) = \begin{cases}
\,\frac{8}{\pi}\,(1-6u^2+6u^3) & {\text{if }} 0 \leqslant u \leqslant \tfrac12,
\\
\,\frac{16}{\pi}\,(1-u)^3 & {\text{if }} \tfrac12 \leqslant u \leqslant 1,
\\
\,0 & {\text{else.}}
\end{cases}
\end{equation}
We repeated the validation with the quartic spline and scaled Gaussian kernels, obtaining consistent results.

We considered a fixed cubic computational domain 80 kpc on a side and used this physical domain for all tests. We then voxelised the particle distribution on a sequence of regular Cartesian grids with $N_{\text{vox}} = N^3$ cells, where $N = 1,2,\ldots,256$ (powers of two). This setup allowed us to study both the behaviour of the algorithms as a function of grid resolution and their consistency under successive refinements of the same domain.

\begin{figure*}
\includegraphics[width=\textwidth]{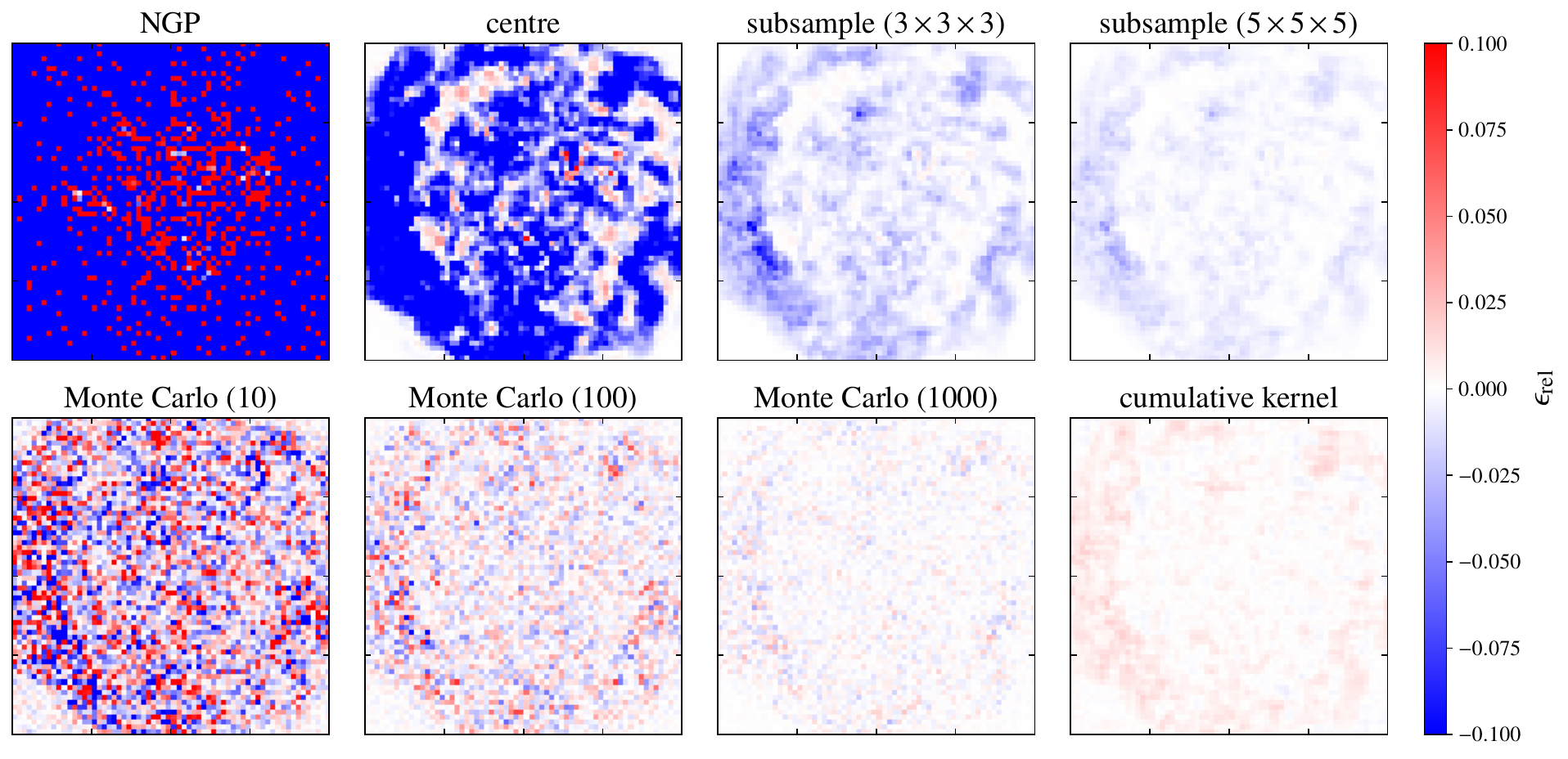}
\caption{Relative errors in the voxelised density field for the different voxelisation methods. Each panel shows $\epsilon_{\text{rel}} = (\rho_{\text{vox}}-\rho_{\text{ref}})/\rho_{\text{ref}}$ in the central plane of the $64^3$ grid, where the reference densities were obtained through direct high-order numerical integration. All panels use the same colour scale. The residuals of the cumulative-kernel method are much smaller than the displayed range.}
\label{NGC628_relerr.fig}
\end{figure*}

\begin{figure*}
\includegraphics[width=\textwidth]{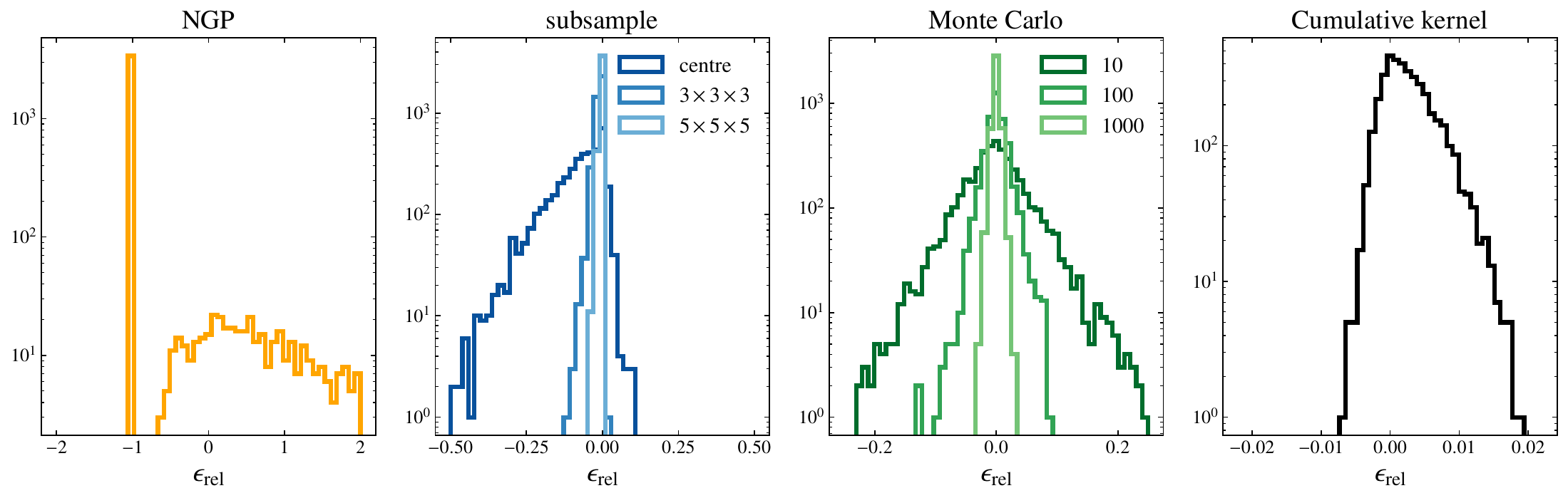}
\caption{Distributions of the relative density errors shown in Fig.~{\ref{NGC628_relerr.fig}}. From left to right: NGP, deterministic sampling, Monte Carlo sampling, and the cumulative-kernel method. Note the different horizontal scales. Increasing the number of sampling points reduces the spread of both the deterministic and Monte Carlo estimates, while the cumulative-kernel errors are substantially smaller than those of the sampling-based methods. The strong peak at $\epsilon_{\text{rel}} = -1$ for NGP corresponds to voxels assigned zero density despite having a non-zero reference density.}
\label{NGC628_histograms.fig}
\end{figure*}

\begin{figure}
\includegraphics[width=0.87\columnwidth]{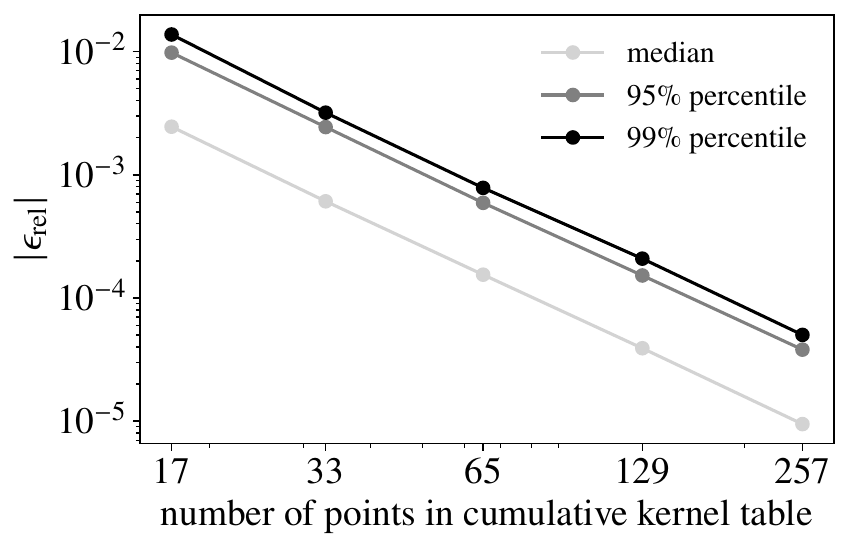}
\caption{Accuracy of the cumulative-kernel voxelisation as a function of lookup-table resolution. Shown are the median, 95th percentile, and 99th percentile of the absolute relative density error $|\epsilon_{\text{rel}}|$ in the central plane of the $64^3$ voxel grid. The errors decrease approximately as $N_{\Phi}^{-2}$, as expected for trilinear interpolation of the smooth cumulative kernel. The lookup table adopted for the calculations in this work contains $129^3$ grid points.}
\label{NGC628_fitstablesize.fig}
\end{figure}

\subsection{Accuracy of the voxelised density field}
\label{Accuracy.sec}

Before considering global mass conservation, consistency under grid refinement, and numerical efficiency of the different voxelisation algorithms, we first investigated how accurately they reproduce the underlying continuous SPH density field. The three diagnostics considered in the subsequent subsections quantify important numerical properties of the voxelisation algorithms but do not directly measure the accuracy of the resulting density distribution. This distinction is particularly relevant for NGP, which conserves mass and is consistent under grid refinement by construction, but completely ignores the spatial distribution described by the smoothing kernel.

To quantify the accuracy of the voxelised density field, we constructed a high-accuracy reference solution by directly integrating the smoothing kernel over each particle--voxel overlap using tensor-product Gauss--Legendre quadrature. We adopted a quadrature order of 64 and verified numerical convergence by repeating the calculation at lower orders. For an order of 32, the median and 99th-percentile relative differences with respect to the order-64 result are $7.2\times10^{-10}$ and $1.4\times10^{-8}$, respectively, demonstrating numerical convergence well beyond the accuracy required here.

For this comparison we considered the central plane of the voxelised density cube. Owing to the orientation of the simulated galaxy within the computational volume, this plane approximately coincides with the midplane of the galaxy and therefore provides a representative view of its density structure. Figure~{\ref{NGC628_density.fig}} shows the density distribution in this plane for grids with $N=256$ and $N=64$. The higher-resolution grid illustrates the structure of the underlying SPH density field, while the $64^3$ grid corresponds to the resolution adopted for the quantitative comparison below.

For each voxelisation method, we calculated the relative density error
\begin{equation}
\epsilon_{\text{rel}}
=
\frac{\rho_{\text{vox}}-\rho_{\text{ref}}}
{\rho_{\text{ref}}},
\end{equation}
where $\rho_{\text{vox}}$ is the voxel density obtained with the method under consideration and $\rho_{\text{ref}}$ is the reference density. Figure~{\ref{NGC628_relerr.fig}} shows the spatial distribution of $\epsilon_{\rm rel}$ in the same central plane of the $64^3$ grid, while Fig.~{\ref{NGC628_histograms.fig}} shows the corresponding relative error distributions.

The differences between the various methods are immediately apparent. NGP provides a very poor representation of the SPH density field, with many voxels containing no mass at all even though the reference density is non-zero. These voxels correspond to the large peak at $\epsilon_{\text{rel}}=-1$ in the left panel of Fig.~{\ref{NGC628_histograms.fig}}. The centre and regular-subsampling methods show a predominantly negative error, indicating a systematic underestimation of the voxel-averaged density. This bias arises because these methods sample the smoothing kernels at a fixed set of positions within each voxel rather than integrating them over the voxel volume. For particles whose kernels are only marginally resolved by the sampling grid, the centrally concentrated kernel profiles are therefore more likely to be undersampled than accurately captured, leading to the observed preference for negative errors. The negative bias remains present for increasing numbers of sampling points, although the magnitude and spread of the errors decrease substantially. In contrast, the Monte Carlo estimates are approximately unbiased and show the expected stochastic fluctuations around zero, with a progressively narrower error distribution for increasing numbers of samples.

The cumulative-kernel method yields by far the smallest density errors. Its residuals are barely visible on the common colour scale in Fig.~{\ref{NGC628_relerr.fig}}, and the corresponding histogram is concentrated within a range of only a few times $10^{-4}$. A small positive bias is visible in the error distribution. This bias originates from the finite resolution of the cumulative-kernel lookup table and the trilinear interpolation used to evaluate it. Because the cumulative kernel is not locally linear, interpolation introduces small systematic errors in the corner values entering the finite-volume estimate. Despite this small interpolation bias, the cumulative-kernel method provides a substantially more accurate representation of the underlying SPH density field than any of the sampling-based methods considered here.

We note that the accuracy advantage of the cumulative-kernel method over the sampling-based approaches is, in principle, shared by other exact finite-volume methods, such as the general polyhedral formulation of \citet{Petkova2018}, which was not included in this comparison. The advantage of the cumulative-kernel algorithm therefore lies not in accuracy, where exact methods are, by construction, comparable. It lies in its combination of simplicity, generality across kernels, and computational efficiency, as demonstrated in the next subsections.

The remaining error of the cumulative-kernel method is dominated by the finite resolution of the cumulative-kernel lookup table and the associated trilinear interpolation. The relative errors shown in Fig.~{\ref{NGC628_histograms.fig}} correspond to the specific choice of a lookup table with $129^3$ grid points. The accuracy can be adjusted by changing the resolution of the lookup table, or equivalently, the number of grid points in each dimension, $N_\Phi$. Figure~{\ref{NGC628_fitstablesize.fig}} illustrates the accuracy of the cumulative-kernel voxelisation method for different lookup table resolutions. The plot shows three measures of the absolute relative density error distribution: the median and the 95th and 99th percentiles. All three error measures decrease approximately as $N_\Phi^{-2}$, consistent with the expected second-order convergence of trilinear interpolation for a sufficiently smooth function \citep{Weiser1988}. For a double-precision lookup table with $129^3$ grid points, the median absolute relative error is approximately $4\times10^{-5}$, which is sufficient for most applications. The storage requirement for a $129^3$ table is about 17~MB. Higher accuracy can be obtained by increasing the resolution of the lookup table, at the cost of increased storage requirements.

\begin{figure}
\includegraphics[width=\columnwidth]{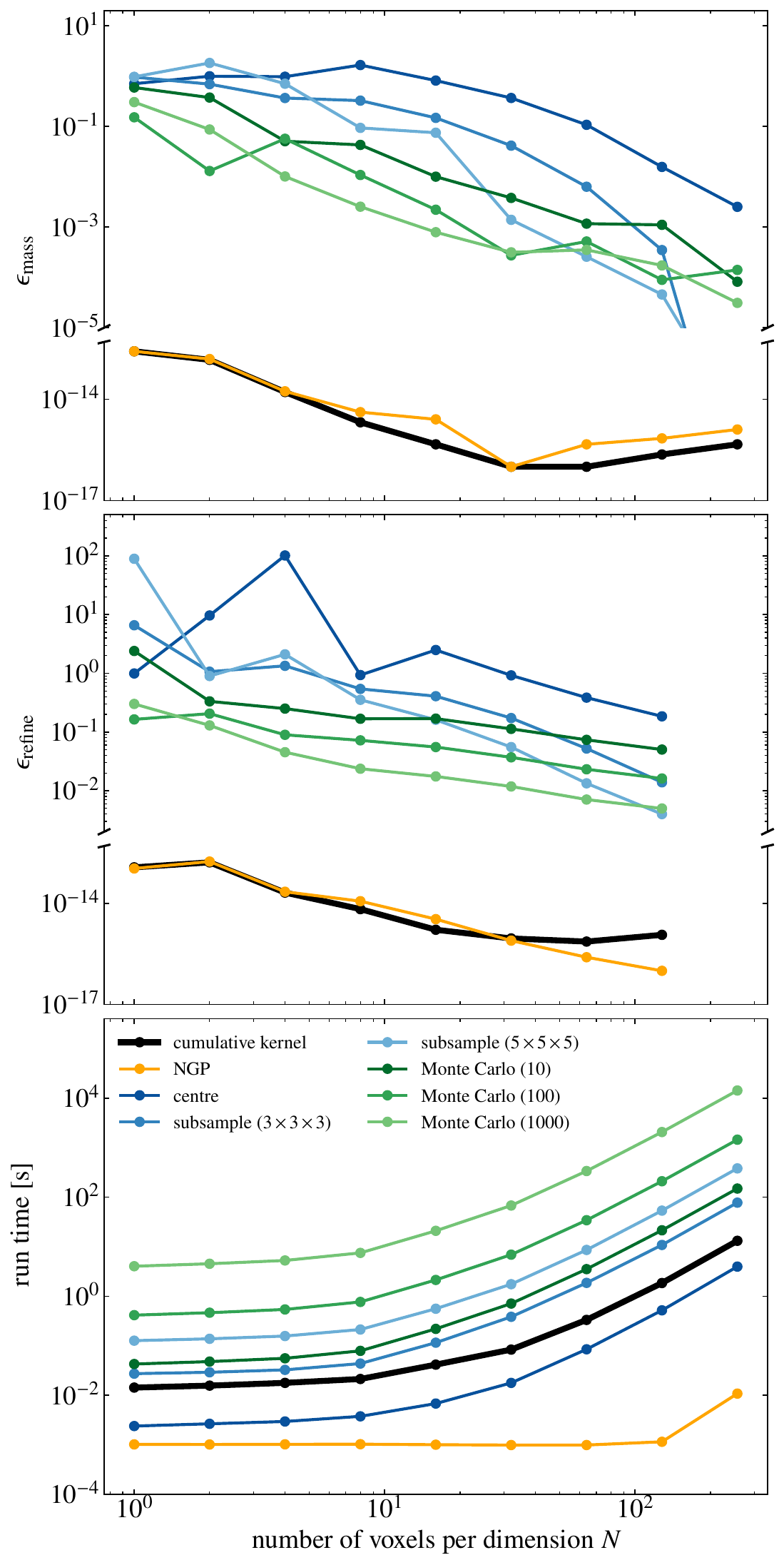}
\caption{Comparison of the voxelisation methods considered in this work. {\it{Top:}} Global mass conservation error as a function of the linear grid resolution $N$. {\it{Middle:}} Refinement consistency error, obtained by comparing the mass assigned to each cell on an $N^3$ grid with the sum of the masses assigned to the corresponding $2\times2\times2$ block of cells on the $(2N)^3$ grid. {\it{Bottom:}} Wall-clock runtime measured in single-thread mode. The broken vertical axes in the upper two panels separate machine-precision errors from the larger errors of the approximate sampling methods. NGP satisfies both the mass-conservation and refinement-consistency criteria by construction but does not represent the spatial distribution encoded by the SPH smoothing kernel.}
\label{Comparison.fig}
\end{figure}

\subsection{Global mass conservation}

We next investigated global mass conservation. We quantified the relative mass conservation error as
\begin{equation}
\epsilon_{\text{mass}} =
\left|
1 - \frac{\sum_i M_i}{\sum_n m_n}
\right|,
\end{equation}
where the numerator represents the total mass contained in the voxel grid and the denominator represents the total mass of the input particle distribution. For this test we only considered the particles whose support was completely contained within the computational domain.

The top panel of Fig.~{\ref{Comparison.fig}} shows the mass conservation error as a function of the grid resolution for all voxelisation methods considered. As expected, NGP conserves mass approximately to machine precision, since each particle is assigned in its entirety to a single voxel. This excellent conservation should not be interpreted as an accurate representation of the SPH density field, however, since NGP ignores the smoothing kernel. The new cumulative-kernel algorithm we present in this paper achieves the same level of accuracy, demonstrating that the finite-volume formulation preserves the total particle mass exactly despite distributing the mass over many voxels. This near-exact conservation is a direct consequence of the finite-difference structure of Eq.~(\ref{Mifinal}) and does not require the individual cumulative-kernel evaluations to be exact. When the mass contributions of a particle are summed over all intersected voxels, cumulative-kernel values associated with internal voxel boundaries enter neighbouring cells with opposite signs and cancel algebraically. Only the contributions at the outer boundary of the particle support remain, so that the full particle mass is recovered up to floating-point round-off. The interpolation errors of the cumulative kernel therefore affect the local distribution of mass among neighbouring voxels, but not the total mass.

All sampling-based methods fail to conserve mass exactly. Centre sampling performs particularly poorly at coarse grid resolutions, where the total grid mass can differ from the particle mass by orders of magnitude. Although the mass conservation error decreases as the grid resolution increases, it remains finite even for the highest resolutions considered. Regular subsampling substantially improves the mass conservation compared to centre sampling, and increasing the number of sampling points leads to a corresponding reduction of the error. Monte Carlo sampling shows a similar behaviour. Increasing the number of random sampling points systematically improves the mass conservation, but the resulting voxelisation remains only approximately mass-conservative. None of the sampling-based approaches reaches the machine-precision-level accuracy obtained by the cumulative-kernel algorithm.

\subsection{Consistency under grid refinement}
\label{GridRefinement.sec}

While global mass conservation is an essential property of any voxelisation algorithm, it does not guarantee that the mass distribution is represented consistently at different grid resolutions. We therefore next considered the stronger requirement of consistency under grid refinement. For every voxelisation calculated on an $N^3$ grid, we compared the mass assigned to each cell with the sum of the masses assigned to the corresponding $2\times2\times2$ block of cells in the voxelisation on the next finer $(2N)^3$ grid. We quantified the corresponding refinement error as
\begin{equation}
\epsilon_{\text{refine}}
=
\frac{\sum_i \left| M_i(N) - \sum_{j=1}^{8} M_{ij}(2N) \right|}
{\sum_i M_i(N)},
\end{equation}
where $M_i(N)$ denotes the mass in cell $i$ of the coarse grid, and $M_{ij}(2N)$ the masses in the eight corresponding cells of the refined grid.

The middle panel of Fig.~{\ref{Comparison.fig}} presents the resulting refinement errors. As expected, the NGP algorithm satisfies this criterion approximately to machine precision, since every particle is assigned entirely to a single voxel, irrespective of the grid resolution. More importantly, the cumulative-kernel algorithm achieves the same level of accuracy. The same cancellation property discussed in the previous subsection also explains the machine-precision consistency under grid refinement. When a coarse cell is subdivided, the cumulative-kernel contributions introduced at the new internal boundaries cancel pairwise when the masses of the daughter cells are summed. Their total therefore reproduces the mass of the parent cell up to floating-point round-off, independently of the interpolation accuracy of the individual cumulative-kernel values.

In contrast, all sampling-based methods exhibit finite refinement errors. Centre sampling performs worst, with errors of order unity at coarse resolutions that decrease only gradually as the grid is refined. Regular subsampling and Monte Carlo sampling both improve upon centre sampling, and increasing the number of sampling points systematically reduces the refinement error. Nevertheless, all sampling-based approaches remain many orders of magnitude less accurate than the cumulative-kernel algorithm. In principle, both regular subsampling and Monte Carlo integration converge towards the exact finite-volume result as the number of sampling points tends to infinity. The residual discrepancies seen here therefore reflect the finite number of sampling points used in the calculations rather than an intrinsic limitation of these methods. Increasing the number of samples further would improve the accuracy, but at a correspondingly increasing computational cost.

The cumulative-kernel algorithm is the only SPH voxelisation method considered here that satisfies this stronger criterion by construction while retaining the spatial information encoded by the smoothing kernel. The resulting voxelisation is therefore resolution-independent: increasing the grid resolution does not alter the represented mass distribution but merely partitions it into smaller voxels. This behaviour follows directly from the finite-volume formulation introduced in Sect.~{\ref{VoxelisationDescription.sec}}. 

\subsection{Numerical efficiency}
\label{Efficiency.sec}

The final diagnostic quantity is the computational efficiency of the different voxelisation algorithms. We measured the wall-clock runtime of each calculation in single-thread mode using the same timing procedure, as described by \citet{Saftly2013} and \citet{Baes2017}. All methods were implemented within the same software framework, such that the measured runtimes reflect only differences in the voxelisation algorithm itself.

The bottom panel of Fig.~{\ref{Comparison.fig}} shows the runtime as a function of the grid resolution. As expected, the computational cost increases with increasing grid resolution for all methods. At the coarsest grid resolutions, the runtime remains nearly independent of the grid resolution. In this regime, the voxel size is much larger than the typical SPH smoothing length, such that each particle contributes to only one or a few voxels. Consequently, the computational cost is dominated by the traversal of the particle list rather than by the number of voxel updates. As the grid resolution increases and the voxel size becomes comparable to or smaller than the smoothing lengths, the number of voxels intersected by each particle grows, leading to the observed increase in runtime.

The centre-sampling algorithm is the fastest SPH-based voxelisation method considered, since it requires only a single kernel evaluation per voxel. Regular subsampling and Monte Carlo sampling become progressively more expensive as the number of sampling points increases, with the computational cost scaling approximately linearly with the number of kernel evaluations per voxel.

The cumulative-kernel algorithm is more expensive than centre sampling but is substantially faster than the higher-order sampling schemes. At the highest grid resolution considered, it is approximately a factor of three slower than centre sampling, but about a factor of six faster than regular $3\times3\times3$ subsampling and more than two orders of magnitude faster than Monte Carlo sampling with 100 sampling points per voxel. Moreover, the relative overhead compared to centre sampling decreases systematically with increasing grid resolution. This behaviour results from the reuse of cumulative-kernel values at corners shared by neighbouring voxels: as the number of voxels intersected by each particle increases, an increasing fraction of the required corner values is shared. Combined with the results of the previous two subsections, these results demonstrate that the cumulative-kernel algorithm achieves mass conservation and consistency under grid refinement approximately to machine precision at a computational cost substantially below that of sampling-based methods that aim for comparable accuracy.

\section{Implementation in {\texttt{SKIRT}}}
\label{SKIRT.sec}

To investigate the applicability of the cumulative-kernel method in a realistic production environment, we implemented it in the general-purpose 3D Monte Carlo radiative transfer code {\texttt{SKIRT}}\footnote{\url{https://skirt.ugent.be/}} \citep{Baes2011b, Camps2015a, Camps2020}. Among its many applications, {\texttt{SKIRT}} can import smoothed particles directly from hydrodynamical simulations. This functionality is widely used to generate synthetic observations of simulated galaxies extracted from cosmological SPH simulations \citep[e.g.][]{Camps2016, Trayford2017, Vijayan2022, Faucher2023, Matsumoto2026, Gebek2026, Lu2026a, Lu2026b, Durrant2026}.

Similar to most radiative transfer codes, {\texttt{SKIRT}} operates with an internal spatial grid on which the medium is discretised. The original density field, whether specified analytically or represented by a set of particles or cells, is mapped onto this spatial grid for subsequent radiative transfer calculations. {\texttt{SKIRT}} is equipped with a suite of spatial grids, including Cartesian, cylindrical, and spherical grids, adaptive octree and binary-tree grids \citep{Saftly2013, Saftly2014}, Voronoi tessellations \citep{Camps2013}, and tetrahedral grids \citep{Lauwers2024}. The code determines the average density in each spatial cell by sampling the input density field at a number of randomly distributed positions within the cell. The default configuration uses 100 sampling points per cell. This procedure is essentially equivalent to the Monte Carlo voxelisation method considered in Sect.~{\ref{ComparisonMethods.sec}}. It provides an increasingly accurate estimate of the cell-averaged density as the number of sampling points is increased, but does not guarantee exact mass conservation or consistency under grid refinement for a finite number of samples.

The cumulative-kernel implementation supports spatial grids whose cells are axis-aligned cuboids. This includes Cartesian grids as well as hierarchical octree and binary-tree grids. The Cartesian grids supported by {\texttt{SKIRT}} are more general than the regular grids considered in the previous sections: the cell boundaries along each coordinate axis can be specified independently, such that the cells need not have identical sizes. This requires no modification of the cumulative-kernel formalism, since Eq.~(\ref{Mifinal}) applies independently to each axis-aligned cuboidal cell, irrespective of its size.

The implementation differs from the standalone voxelisation algorithm described in Sect.~{\ref{Implementation.sec}} in the order in which particles and cells are processed. For a regular Cartesian grid, it is most efficient to iterate over the particles and determine the voxels intersected by each particle, as locating these voxels is straightforward. Moreover, cumulative-kernel values at corners shared by adjacent voxels can be calculated only once and reused. In {\texttt{SKIRT}}, the iteration is instead performed over the spatial cells. For each cell, the code first identifies the particles whose smoothing volumes potentially overlap the cell and subsequently uses the cumulative kernel to determine the mass contributed by each of these particles. This cell-oriented approach fits naturally within the more general {\texttt{SKIRT}} architecture and can be applied in exactly the same manner to Cartesian, octree, and binary-tree grids.

The cumulative-kernel method is particularly useful for the construction of hierarchical grids. In an octree or binary-tree grid, cells are recursively subdivided according to a refinement criterion that is usually based on the mass contained in the cell \citep{Saftly2013, Saftly2014}. The cumulative-kernel method is therefore used not only to determine the final cell densities but also during construction of the grid to evaluate the mass-based subdivision criterion. This application directly benefits from the consistency under grid refinement demonstrated in Sect.~{\ref{GridRefinement.sec}}: the mass assigned to a parent cell is, apart from numerical round-off, identical to the sum of the masses subsequently assigned to its daughter cells. As a result, mass-based refinement becomes independent of the current grid level, leading to a more robust and predictable construction of adaptive grids.

Besides improving the numerical properties of the density assignment, the new implementation also provides a substantial performance gain. In initial performance tests with {\texttt{SKIRT}}, replacing the default Monte Carlo estimate based on 100 random samples per cell by the cumulative-kernel method accelerated the calculation of the cell densities by approximately a factor of five. For hierarchical grids, the complete tree construction was accelerated by approximately a factor of four. These factors depend on the particle distribution, grid structure, and computing system.

These speed-ups are smaller than those obtained with the standalone implementation discussed in Sect.~{\ref{Efficiency.sec}}. This is primarily a consequence of the more general cell-oriented architecture in {\texttt{SKIRT}}. For each cell, {\texttt{SKIRT}} needs to identify the potentially overlapping particles, and cumulative-kernel values at corners shared by neighbouring cells cannot be reused. In addition, particularly during hierarchical grid construction, density evaluation represents only part of the total computational work. Nevertheless, the substantial acceleration obtained in this more complex setting demonstrates that the cumulative-kernel approach provides significant practical benefits in production radiative transfer calculations.

\section{Conclusions}
\label{Conclusions.sec}

We presented a new finite-volume algorithm for the voxelisation of SPH particle distributions onto regular Cartesian grids. The method is based on a cumulative kernel function, which transforms the 3D volume integral associated with every particle--voxel overlap into a linear combination of cumulative-kernel values at the eight voxel corners. Values at corners shared by neighbouring voxels can be evaluated once and reused, substantially reducing the computational cost. In contrast to previous finite-volume formulations, the present approach requires neither kernel-specific analytical antiderivatives nor geometric operations on polyhedral cells, making the implementation both simple and independent of the adopted smoothing kernel.

We argued that a suitable SPH voxelisation algorithm should satisfy four key requirements: an accurate representation of the underlying SPH density field, global mass conservation, consistency under grid refinement, and computational efficiency. Using an SPH simulation of an isolated disc galaxy, we compared the cumulative-kernel algorithm to NGP assignment, centre sampling, regular subsampling, and Monte Carlo integration. Direct high-order numerical integration was used to establish an independent reference for the voxel densities. The cumulative-kernel method reproduces these reference densities substantially more accurately than the sampling-based approaches, while conserving mass and remaining consistent under grid refinement approximately to machine precision. For the adopted $129^3$ lookup table, the median absolute relative density error is about $4\times10^{-5}$, and the error can be reduced systematically by increasing the lookup-table resolution. We emphasise that this accuracy figure, unlike the mass-conservation and refinement-consistency results above, is bounded by the lookup-table resolution rather than being exact to machine precision. The numerical experiments demonstrate that the cumulative-kernel algorithm is the only SPH voxelisation method considered that satisfies both the mass-conservation and refinement-consistency criteria approximately to machine precision while also being substantially more efficient than the other methods that account for the finite extent of the smoothing kernel.

The practical applicability of the method is demonstrated by its implementation in the {\texttt{SKIRT}} radiative transfer code. The cumulative-kernel approach can be used not only with Cartesian grids with arbitrarily spaced cell boundaries, but also with hierarchical octree and binary-tree grids. In the latter case, it is used both to determine the final cell densities and to evaluate mass-based subdivision criteria during construction of the adaptive grid. Initial performance tests show substantial computational gains compared to the default Monte Carlo approach based on 100 sampling points per cell, with density calculations accelerated by approximately a factor of five and hierarchical tree construction by approximately a factor of four.

A further advantage of the cumulative-kernel formulation is that it is not tied to a particular smoothing kernel. The kernel dependence is entirely contained in the cumulative-kernel lookup table, while the voxelisation algorithm itself remains unchanged. Such lookup tables can therefore be constructed once for any desired spherically symmetric compact smoothing kernel and subsequently used with the same voxelisation machinery. We provide precomputed cumulative kernels for several commonly used SPH kernels: the cubic spline \citep{Monaghan1985}, the quartic spline \citep{Dehnen2012}, and the scaled Gaussian kernel \citep{Altay2013}. Lookup tables for other kernels \citep[e.g.][]{Liu2003, Hongbin2005, Dehnen2012} can readily be generated using the code supplied with this work.

The cumulative-kernel approach is also not restricted to density. Once the finite-volume contribution of each particle to a cell has been determined, the same weights can be used to construct mass-weighted cell averages of other particle-carried quantities, such as velocity, temperature, or metallicity. The method can therefore be applied more generally to the transfer of SPH particle properties onto spatial grids.

Finally, the cumulative-kernel approach is not restricted to 3D voxelisation. With minimal adaptation, the same principle can be applied to related 2D problems. For example, it can be used to determine the mass distribution in a pixelised planar cross-section through an SPH particle distribution, or to construct a pixelised projection of the mass distribution along one of the coordinate axes \citep[e.g.][]{Borrow2021}. In both cases, the required integrals can be expressed in terms of appropriately defined cumulative kernels, retaining the deterministic finite-volume character of the method. These extensions may be useful for constructing projected maps or cross-sections directly from particle-based simulations without resorting to spatial sampling.

\begin{acknowledgements}
MB and AN gratefully acknowledge the support of the Belgian Federal Science Policy Office (BELSPO)
for the provision of financial support in the framework of the PRODEX Programme of the European Space Agency (ESA) under contract numbers 4000143347 and 4000147090. KM is a postdoctoral fellow of the Flemish Fund for Scientific Research (FWO-Vlaanderen, grant 12AJQ26N). 
\end{acknowledgements}

\bibliography{mybib}

\end{document}